\documentclass[final,3p,times,twocolumn]{elsarticle}

\usepackage{graphicx}
\usepackage{dcolumn}
\usepackage{bm}
\usepackage{epsf}
\usepackage{amsmath}
\usepackage{amssymb}
\usepackage{color}
\usepackage[colorlinks=true,citecolor=blue,linkcolor=blue]{hyperref}

\newcommand{\bra}[1]{\langle #1|}
\newcommand{\ket}[1]{|#1\rangle}

\newcommand{\tr}[1]{\mathrm{tr}\left\{#1\right\}}

\newcommand{\la}{\left\langle}
\newcommand{\ra}{\right\rangle}
\newcommand{\pd}{\partial}
\newcommand{\de}[1]{\delta\left(#1\right)}
\newcommand{\td}{\mathrm{d}}

\newcommand{\e}[1]{\exp{\left(#1\right)}}

\newcommand{\bla}{bla\\bla\\bla\\bla\\bla}
\newcommand{\PR}{Phys. Rev.}
\newcommand{\PRA}{Phys. Rev. A }

\newcommand{\PRE}{Phys. Rev. E }
\newcommand{\PRL}{Phys. Rev. Lett. }
\newcommand{\PRX}{Phys. Rev. X }

\newcommand{\RMP}{Rev. Mod. Phys. }

\newcommand{\mc}[1]{\mathcal{#1}}

\DeclareMathOperator*{\sumint}{%
\mathchoice%
  {\ooalign{$\displaystyle\sum$\cr\hidewidth$\displaystyle\int$\hidewidth\cr}}
  {\ooalign{\raisebox{.14\height}{\scalebox{.7}{$\textstyle\sum$}}\cr\hidewidth$\textstyle\int$\hidewidth\cr}}
  {\ooalign{\raisebox{.2\height}{\scalebox{.6}{$\scriptstyle\sum$}}\cr$\scriptstyle\int$\cr}}
  {\ooalign{\raisebox{.2\height}{\scalebox{.6}{$\scriptstyle\sum$}}\cr$\scriptstyle\int$\cr}}
}

\begin{document}

\begin{frontmatter}

\title{Quantum work distribution for a driven diatomic molecule}
\author[BIO]{Alison Leonard}
\ead{aleonar2@terpmail.umd.edu}
\author[UMD,LANL]{Sebastian Deffner}
\ead{sdeffner@lanl.gov}
\address[BIO]{Biophysics Program, Institute For Physical Science and Technology, University of Maryland, College Park, Maryland 20742, USA}
\address[UMD]{Department of Chemistry and Biochemistry and Institute for Physical Science and Technology, University of Maryland,
College Park, Maryland 20742, USA}
\address[LANL]{Theoretical Division and Center for Nonlinear Studies, Los Alamos National Laboratory, Los Alamos, New Mexico 87545, USA}
\date{\today}

\begin{abstract}
We compute the quantum work distribution for a driven Morse oscillator. To this end, we solve the time-dependent dynamics for a scale-invariant process, from which the exact expressions for the transition probabilities are found. Special emphasis is put on the contributions to the work distribution from discrete (bound) and continuous (scattering) parts of the spectrum. The analysis is concluded by comparing the work distribution for the exact Morse potential and the one resulting from a harmonic approximation. 
\end{abstract}


\end{frontmatter}

\section{\label{sec:intro}Introduction}

If definitions of quantum work are to be experimentally relevant, we have to develop theoretical frameworks that can predict the outcomes of actual experiments. For isolated quantum systems that evolve under unitary dynamics, the so-called two-time energy measurement approach \cite{kurchan_2000,tasaki_2000,talkner_2007,campisi_2011} has been proven to be powerful and practical. The first law of thermodynamics states that the change of internal energy, $\Delta E=\la H_{t_f}\ra-\la H_0\ra$ with $H_t$ being the time-dependent Hamiltonian, during a process of duration $t_f$, can be separated into work, $\la W\ra$, and heat, $\la Q\ra$, and we have $\Delta E=\la W\ra+\la Q\ra$. The angular brackets denote an average over an ensemble of realizations of the same process. In particular, we have $\la W\ra=\int\td W\, \mc{P}(W)$, where $\mc{P}(W)$ is the probability distribution of the work.

For isolated systems, no heat is exchanged with any environment and all changes of the internal energy are identified as the work performed by the system under study, $\Delta E=\la W\ra$. Thus, quantum work can be determined by an initial and a final projective energy measurement on the quantum system, which has been successfully applied, for instance, theoretically in Refs.~\cite{deffner_2008,talkner_2008,talkner_2008a,deffner_2010_prl,deffner_2010,deffner_2011,quan_2012,fusco_2014} and  experimentally in Refs.~\cite{huber_2008,mazzola_2013a,dorner_2013,batahlao_2014}. It also has been shown that the two-time energy measurement approach is of practical relevance. In particular, experimentally realizable nanoengines with single ions as working medium have been proposed  \cite{abah_2012,abah_2013,rossnagel_2013}. 

Nevertheless, previous studies have been mostly restricted to simple systems as, for instance, the driven harmonic oscillator \cite{deffner_2008,talkner_2008,deffner_2010,campo_2014} or square well potentials \cite{quan_2012}. The reason is that in order to compute the probability distribution, $\mc{P}(W)$, the corresponding time-dependent Schr\"odinger equation has to be solved.

Only recently it has been recognized that these kinds of problems are greatly simplified for so-called scale invariant processes \cite{deffner_2014,zheng_2014}. Scale-invariant driving is generated by transformations of $H_t$ for which the density profile (and all correlations in real space) is preserved up to scaling and translation. In this case a solution of the time-dependent Schr\"odinger equation is given by the instantaneous eigenfunctions of $H_t$ multiplied by a phase \cite{berry_1984}.

\begin{figure}
\includegraphics[width = .45\textwidth]{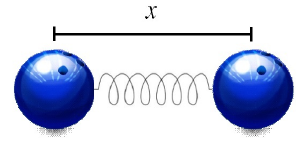}
\caption{\label{fig:spring} \textbf{Illustration of the model system:} The Morse potential mimics the electrostatic potential of a diatomic bond of the length $x$.}
\end{figure}

In the present paper we apply this method to compute the quantum work distribution for an important system in chemical physics, namely, the driven Morse oscillator.  The Morse oscillator is a well-studied system under continued interest \cite{alhassid_1983,alhassid_1984,benjamin_1985,dahl_1988,frank_2000,dong_2002,rawitscher_2002,demirplak_2005,angelova_2008,dittrich_2010,mccoy_2011} that mimics the covalent bond in a diatomic molecule \cite{morse_1929,flugge_1971}; see also Fig.~\ref{fig:spring}.
Generally, analytical solutions are only known for the time-independent case \cite{flugge_1971}, whereas the time-dependent case has only been solved for specific situations \cite{kondo_1988,lima_2006}. An additional complication is that the eigenvalue spectrum has a discrete part, corresponding to bound states, and a continuum of scattering states. This complexity of the spectrum poses a theoretically interesting problem -- ubiquitous in molecular physics and realistic systems -- and it will be reflected in the shape of the resulting work density.

In addition, this paper is also of pedagogical value; the same approach could readily be applied to any quantum system for which an exact or approximate Schr\"odinger equation is known. In the following, we will show how to generally construct the time-dependent solutions for scale-invariant driving from which the quantum work distribution is determined.

The paper is organized as follows: In Sec.~\ref{sec:def} we collect definitions and establish notation. Section~\ref{sec:scale} is dedicated to solving the time-dependent Morse oscillator for scale invariant driving, before we finally analyze the quantum work distribution in Sec.~\ref{sec:work}. In the analysis we will focus on the contribution of bound and scattering states, and we will compare the exact solution to a harmonic approximation. We conclude the discussion with a few remarks in Sec.~\ref{sec:con}

\section{\label{sec:def}Definitions and notation}

\subsection{Two-time energy measurement approach}

Consider an isolated quantum system with time-dependent Schr\"odinger equation
\begin{equation}
\label{eq01}
i\hbar\,\pd_t \ket{\psi_t}=H_t\,\ket{\psi_t}\,.
\end{equation}
In the following we are interested in describing thermodynamic processes that are induced by varying an external control parameter $\lambda_t$ during time $t_f$, so that $H_t=H(\lambda_t)$. Within the two-time energy measurement approach quantum work is determined by the following protocol: At initial time $t=0$ an projective energy measurement is performed on the system; then the system is let to evolve under the time-dependent Schr\"odinger equation \eqref{eq01}, before a second projective energy measurement is performed at $t=t_f$. For a single realization of this protocol the work is given by
\begin{equation}
\label{eq02}
W\left[\ket{n_0(\lambda_0)}; \ket{n_f(\lambda_{t_f})}\right]=E_{n_f}(\lambda_{t_f})-E_{n_0}(\lambda_0)\,,
\end{equation}
where $\ket{n_0}$ is the initial eigenstate with eigenenergy $E_{n_0}(\alpha_0)$ and $\ket{n_f} $ with $E_{n_f}(\alpha_{t_f}) $ denotes the final state.

The distribution of work values is then given by averaging over an ensemble of realizations of the same process,
\begin{equation}
\label{eq03}
\mc{P}(W)=\la \de{W-W\left[\ket{n_0(\lambda_0)}; \ket{n_f(\lambda_{t_f})}\right]} \ra \,,
\end{equation}
which can be rewritten as \cite{kafri_2012,deffner_2013}
\begin{equation}
\label{eq04}
\mathcal{P}(W)=\sumint_{n_0,n_f} \de{W-W\left[\ket{n_0};\ket{n_f}\right]}\,p\left(\ket{n_0}\rightarrow\ket{n_f}\right).
\end{equation}
In the latter equation the symbol $\sumint$ denotes that we have to sum over the discrete part of the eigenvalues spectrum and integrate over the continuous part. Therefore, for systems with spectra that have both contributions the work distribution will have a continuous part and delta-peaks, a fact that we will analyze more carefully for the Morse oscillator, shortly. 

Further, $p\left(\ket{n_0}\rightarrow\ket{n_f}\right)$ denotes the probability to observe a specific transition $\ket{n_0}\rightarrow\ket{n_f} $. This probability is given by \cite{kafri_2012},
\begin{equation}
\label{eq05}
p\left(\ket{n_0}\rightarrow\ket{n_f}\right)=\tr{\Pi_{\nu_f}\, U_{t_f}\, \Pi_{n_0}\,\rho_0\,\Pi_{n_0}\, U_{t_f}^\dagger}\,,
\end{equation}
where $\rho_0$ is the initial density operator of the system and $U_{t_f}$ is the unitary time evolution operator, $U_{t_f}=\mc{T}_> \e{-i/\hbar\,\int_0^{t_f}\td t\,H_t}$. Finally, $\Pi_n$ denotes the projector into the space spanned by the $n$th eigenstate. For Hamiltonians with non-degenerate spectra we simply have $\Pi_n=\ket{n}\bra{n}$.

From Eqs.~\eqref{eq02}-\eqref{eq05} we see that, in order to compute $\mc{P}(W)$, we have to determine the initial state, $\rho_0$, the instantaneous eigenvalues, $E_n(\lambda_t)$, and the solution of the dynamics, $U_t$. 

Generally, the initial state can be chosen according to the physical situation and the energy eigenvalues are given by the time-independent problem. To determine the transition probabilities \eqref{eq05}, however, the time-dependent Schr\"odinger equation \eqref{eq01} has to be solved.

\subsection{Time-independent solution of the Morse oscillator}

In the following we will study the quantum work distribution \eqref{eq04} for the time-dependent Morse oscillator. Before we find a solution to the corresponding Schr\"odinger equation \eqref{eq01}, let us briefly summarize the time-independent solution.

The Morse potential \cite{morse_1929} describes approximately the electrostatic interaction constituting the covalent bond in a diatomic molecule. It can be written as
\begin{equation}
\label{eq06}
V(x) = V_0 \left(\e{-2Bx} - 2\,\e{-Bx}\right)\,.
\end{equation}
where $V_0$ is the depth of the potential well, and $B$ determines its width.  Finally, $x$ denotes the nuclear separation of the atoms in the bond; see also Fig.~\ref{fig:spring}.  For an actual diatomic molecule, $B$ depends on the molecules' reduced mass $\mu$, the potential depth $V_0$ of the bond, and the vibrational constant $\omega_e$ associated with the bond, which can be determined experimentally \cite{kaplan_2003}. In particular, with $c$ being the speed of light we have 
\begin{equation}
\label{eq07}
 B = 2\pi c\, \omega_e\,\sqrt{\frac{\mu}{2 V_0}}\,.
\end{equation}
In Fig.~\ref{fig:spring} we sketch the diatomic bond as a spring connecting the two nuclei. Mathematically, this corresponds to a harmonic expansion of the actual potential,
\begin{equation}
\label{eq08}
V(x) \simeq -V_0+ V_0 \,B^2\, x^2+\mc{O}(x^3)\,.
\end{equation} 
\begin{figure}
\includegraphics[width = .46\textwidth]{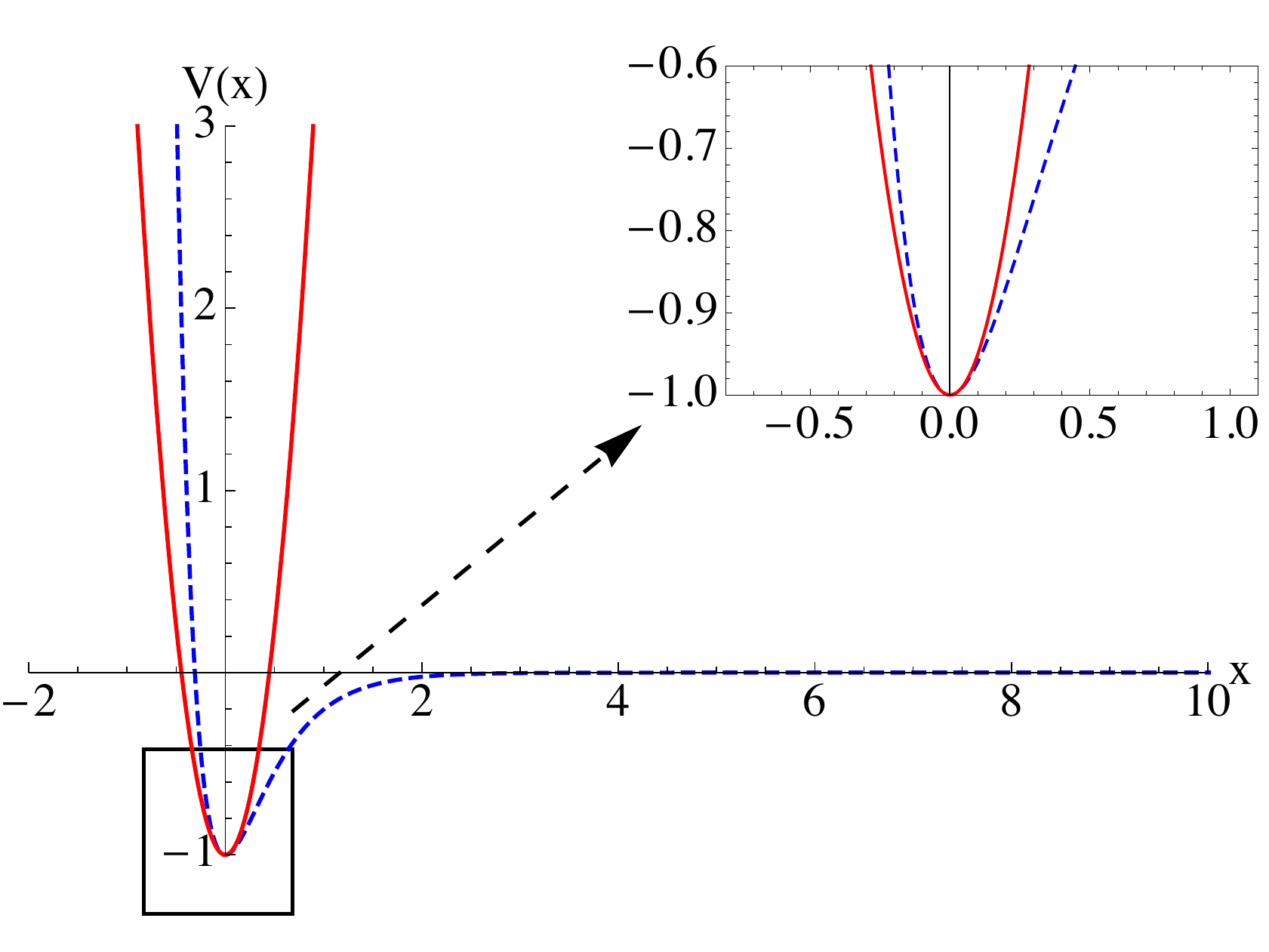}
\caption{\label{fig:potential} \textbf{Illustration of the harmonic approximation:} Morse potential $V(x)$ \eqref{eq06} (blue, dashed line) together with its harmonic approximation \eqref{eq08} (red, solid line)  for $V_0=1$, $\mu=1$, $k=10$ and $\hbar=1/2\pi$.}
\end{figure}
Thus, for small values of $x$ the Morse potential \eqref{eq06} can be approximated by an harmonic oscillator with spring constant $k = 2\,V_0 B^2$, and $ V(x)\simeq -V_0+k/2\,x^2$,  what is illustrated in Fig.~\ref{fig:potential}.

We also clearly observe that the Morse potential significantly deviates from the harmonic approximation as the distance from $x = 0$ increases. This allows to model bond weakening and dissociation with diverging nuclear separation. The consequence of this property together with the presence of scattering states at higher energies, introduces interesting features into the work distribution, as we will discuss shortly in Sec.~\ref{sec:work}.

The time-independent Morse oscillator can be solved exactly \cite{flugge_1971}. For energies less than zero the solution corresponds to bound states, which can be expressed as \cite{lima_2006}
\begin{equation}
\label{eq09} 
\phi(z,\nu)= N_b(\nu)\, z^{A - \nu}\, \e{-z/2}\, M\left(-\nu, 2A-2\nu +1, z\right)\,,
\end{equation}
where $M(a,b,c)$ is the Kummer function of the first kind \cite{abramowitz_1964}. We further introduced the new variable $z = (2A + 1)\,\e{-Bx} $ and the parameter $A$ with $(A + 1/2)^2 = 2\mu V_0/\hbar^2 B^2$. Finally, $\nu$ is the energy quantum number ranging from $0$ to the integer part of $A$. The prefactor $N_b(\nu)$ is a normalization constant that can be written as,
\begin{equation}
\label{eq10}
N_b(\nu) = \left[\frac{B\,(2A - 2\nu)\,\Gamma(2A - \nu +1)}{\nu!\,\Gamma(2A-2\nu+1)^2}\right]^{1/2}\,,
\end{equation}
where $\Gamma(\cdot)$ denotes the Euler gamma function \cite{abramowitz_1964}.

The eigenenergies associated with the energy eigenstates \eqref{eq09} read,
\begin{equation}
\label{eq11}
E(\nu) = -\hbar^2\,(A-\nu)^2 B^2/2\mu
\end{equation}
which are negative and whose difference decreases quadratically in $\nu$. For the later analysis it will prove convenient to work with the eigenstates being expressed in terms of Kummer functions. However, it is worth mentioning that an equivalent expression in terms of Laguerre polynomials can be found \cite{dong_2002}.

\begin{figure}
\includegraphics[width=.46\textwidth]{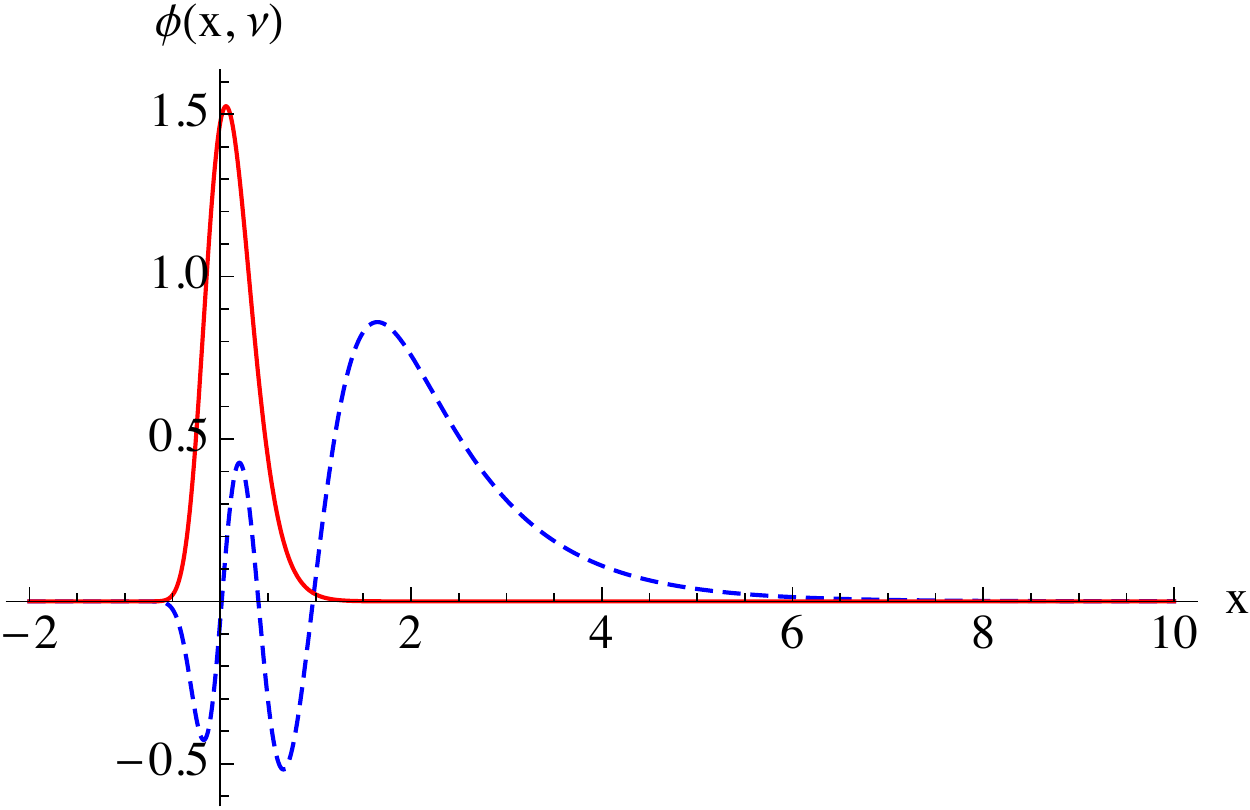}
\caption{\label{fig:bound}\textbf{Bound states of the Morse potential:} Lowest, $\nu=0$, (red, solid line) and highest, $\nu=8$, (blue, dashed line) bound  state of the Morse potential depicted in Fig.~\ref{fig:potential} with $V_0=1$, $\mu=1$, $k=10$ and $\hbar=1/2\pi$.}
\end{figure}

In Fig.~\ref{fig:bound} we plot lowest and highest bound states for the Morse potential illustrated in Fig.~\ref{fig:potential}. We observe again that for small energies the ground state is similar to the one of the harmonic oscillator, whereas the highly excited states exhibit a much richer shape.

The scattering states can be represented similarly \cite{lima_2006}. We have for the eigenfunctions,
\begin{equation}
\label{eq12}
 \phi(z,\kappa) = N_s(\kappa)\,z^{i\kappa}\,\e{-z/2}\,U(-p-i\kappa, 1-2i\kappa, z)\,,
\end{equation}
where $U(a,b,c)$ is now the Kummer function of the second kind \cite{abramowitz_1964} and $N_s(\kappa)$ is again a normalization, that reads here,
\begin{equation}
\label{eq13}
N_s(\kappa) = \frac{\left|\Gamma(-A-i\kappa)\right|}{\pi} \sqrt{B\kappa\sinh(2\pi\kappa)}\,.
\end{equation}
In contrast to the bound state the scattering states correspond to a continuous part of the spectrum, which is characterized by the positive and real quantum number $\kappa$. This quantum number can be associated with the momentum, $p=\hbar \kappa$, and the eigenenergies become accordingly,
\begin{equation}
\label{eq14}
E(\kappa)=\left(\hbar\kappa\right)^2 B^2/2\mu\,\,.
\end{equation}
In complete analogy to the bound states also the scattering states can be expressed in terms of Laguerre polynomials \cite{lima_2006}.

\begin{figure}
\includegraphics[width = .46\textwidth]{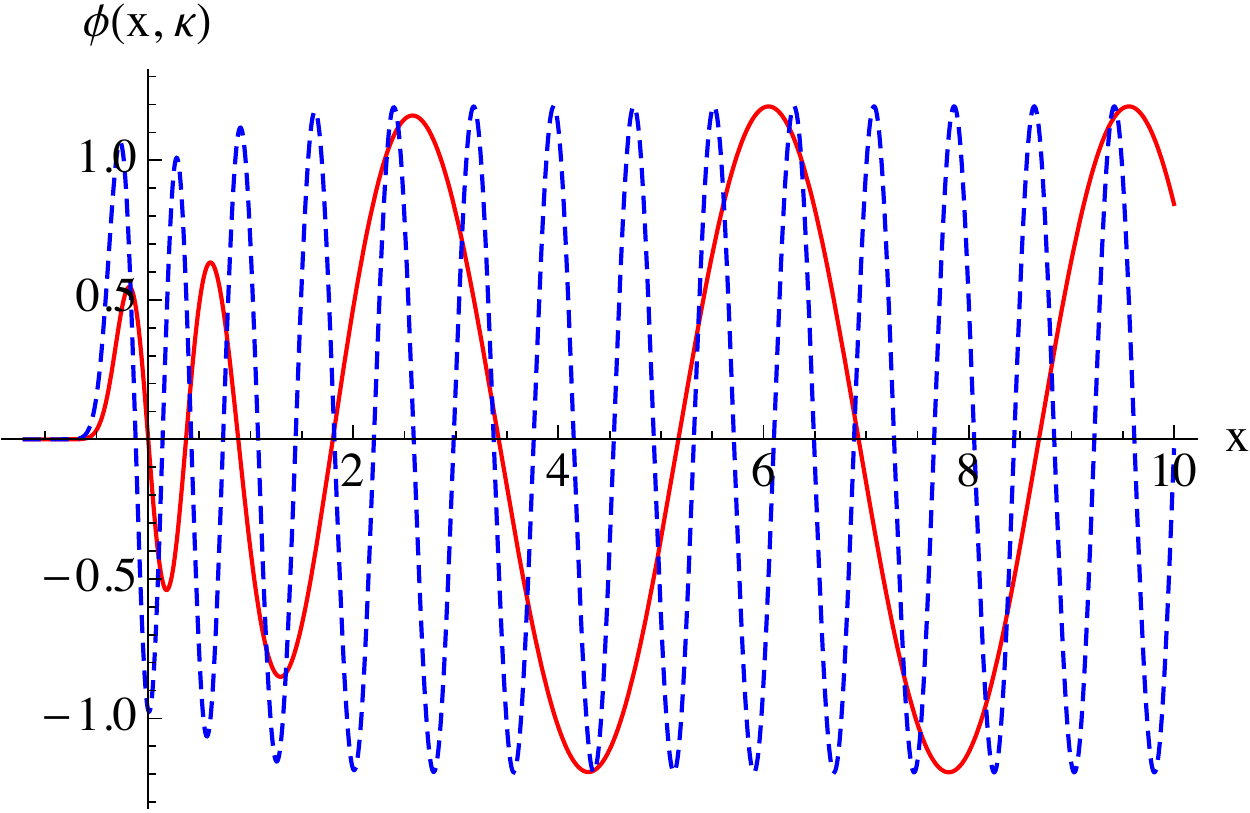}
\caption{\label{fig:scat}\textbf{Scattering states of the Morse potential.} Two scattering states for $\kappa=0.8$ (red, solid line) and $\kappa=3.6$ (blue, dashed line)  of the Morse potential depicted in Fig.~\ref{fig:potential} with $V_0=1$, $\mu=1$, $k=10$ and $\hbar=1/2\pi$.}
\end{figure}

In Fig.~\ref{fig:scat} we plot two scattering states of the Morse potential of Fig.~\ref{fig:potential}. We observe that the scattering states behave like free waves for large diatomic separation, whereas they exhibit interesting structure above the potential well. We will rediscover these features in the shape of the quantum work distribution shortly. 

\section{\label{sec:scale}Scale-invariant driving of the Morse oscillator}

Equipped with the time-independent eigenstates and eigenenergies we can now proceed to solve the time-dependent problem. Generally, it is hardly feasible to solve the time-dependent Schr\"odinger \eqref{eq01} equation analytically. However, it has recently been recognized in the context of so-called shortcuts to adiabaticity that for scale-invariant processes the situation greatly simplifies \cite{jarzynski_2013,campo_2013,deffner_2014}. Scale-invariant driving refers to transformations of the system Hamiltonian associated with a set of external control parameters $\lambda_t$ that can be absorbed by scaling of the coordinates to rewrite the transformed Hamiltonian in its original form up to a multiplicative factor. In particular, for driving protocols of the form
\begin{equation}
\label{eq15}
V(x,\lambda_t)=1/\lambda_t^2\,\, V\left(x/\lambda_t\right)\,,
\end{equation}
an exact solution for the time-dependent equation \eqref{eq01} can be constructed from the time-independent eigenfunctions of the potential \cite{berry_1984}. Berry and Klein showed \cite{berry_1984} that a canonical transformation can be performed to scale the coordinates of position and time such that the shape of a trajectory is unaffected by the presence of a non-conservative force. For quantum systems an analogous canonical transformation of the coordinates yields wave functions to maintain their same shape when they are subjected to a time-dependent Hamiltonian, provided that the rate of change is at most quadratic in time.

\begin{figure}
\includegraphics[width = .46\textwidth]{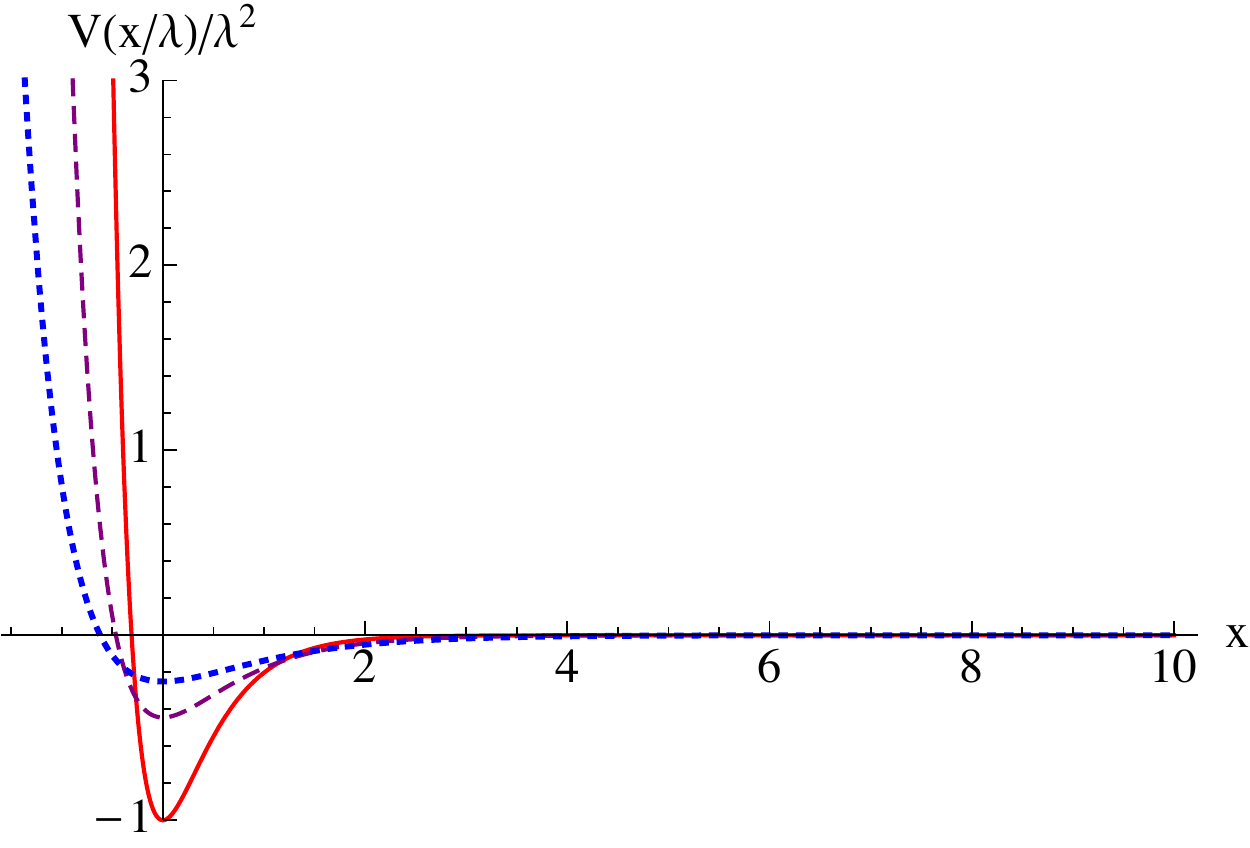}
\caption{\label{fig:scaling}\textbf{Morse potential under scale-invariant transformation:} Morse potential \eqref{eq16} for $\lambda=1$ (red, sold line), $\lambda=1.5$ (purple, dashed line), and $\lambda=2$ (blue, dotted line) with $V_0=1$, $\mu=1$, $k=10$ and $\hbar=1/2\pi$.}
\end{figure}

For our specific case we, therefore, continue by analyzing the scale-invariantly stretched potential
\begin{equation}
\label{eq16}
V(x,\lambda_t) = \frac{V_0}{\lambda_t^2}\, \left(\e{\frac{-2Bx}{\lambda_t}}-2\,\e{\frac{-Bx}{\lambda_t}}\right)\,.
\end{equation} 
In Fig.~\ref{fig:scaling} we plot a few realizations of the latter potential for different values of $\lambda$.

For the sake of simplicity we will further assume that $\lambda_t$ is a linear function in time,
\begin{equation}
\label{eq17}
\lambda_t =1 +\alpha  \,t\,,
\end{equation}
which corresponds physically to a diatomic molecule that is stretched at constant rate, $\alpha$.

Then, a solution of the time-dependent Schr\"ondinger equation \eqref{eq01} with \eqref{eq16}, \eqref{eq17},  and $\alpha=1$ is given by \cite{berry_1984}
\begin{equation}
\label{eq18}
\psi\left(x,t, n\right) = \frac{1}{\sqrt{\lambda_t}}\,\phi\left(\frac{x}{\lambda_t},n\right)\, \e{-\frac{iE_\nu \tau_t}{\hbar}}\,\e{i\,\varphi(x,t)}
\end{equation}
where $\phi(x,n)$ are the time-independent eigenfunctions \eqref{eq09} and \eqref{eq12} with their respective eigenenergies \eqref{eq11} and \eqref{eq14} and $n$ is a generic quantum number referring to either discrete $\nu$ or continuous $\kappa$. The time like variable $\tau_t$ is given by
\begin{equation}
\label{eq19}
\tau_t=\int_0^t\frac{\td t}{\lambda_t^2}
\end{equation}
and $\varphi$ is a phase factor,
\begin{equation}
\label{eq20}
\varphi(x,t) = \frac{\mu}{2\hbar}\,\lambda_t\dot{\lambda_t}\,x^2\,.
\end{equation}
A general solution is then given by
\begin{equation}
\label{eq21}
\Psi(x,t) = \sumint_{n} c_{n_0, n}\, \psi(x,t,n)
\end{equation}
where the coefficients are determined from the  initial condition,
\begin{equation}
\label{eq22}
 c_{n_0, n} = \frac{1}{\lambda_0} \int \td x\,\ \e{-i\, \varphi(x,0)}\, \phi\left(\frac{x}{\lambda_0}, n_0\right)\,\phi\left(\frac{x}{\lambda_0},n\right)\,,
\end{equation}
from which we obtain the transition probabilities \eqref{eq05}.

So far the analysis is exact and completely general. For the following we will choose a specific initial state for which the analysis and the presentation of the results greatly reduces. Let us choose for the sake of simplicity, but without loss of generality that the system is initially prepared in its ground state,
\begin{equation}
\label{eq23}
\psi(x,0)=\phi(x,\nu=0)
\end{equation}
Then the transition probability \eqref{eq05} becomes
\begin{equation}
\label{eq24}
p\left(\ket{n_0}\rightarrow\ket{n_f}\right)=\delta_{n_0,\nu=0}\,\left|\int\td x\,\Psi(x,t_f)\,\frac{\phi(x/\lambda_{t_f},n_f)}{\sqrt{\lambda(t_f)}}\right|^2\,,
\end{equation}
where the time dependent solution reduces to
\begin{equation}
\label{eq25}
\Psi(x,t) = \sumint_{n} c_{\nu=0, n}\, \psi(x,t,n)\,.
\end{equation}
It is worth emphasizing at this point that the time-dependent solution \eqref{eq25} and, hence, the transition probability \eqref{eq24} separates into two contributions: a discrete part stemming from the bound part of the spectrum, and a continuous part contributed by the scattering states. Therefore, also the quantum work distribution \eqref{eq04} has two contributions, discrete and continuous, a feature that has not been seen in previous analyses \cite{deffner_2008,talkner_2008,talkner_2008a,deffner_2010,quan_2012}.

Before we continue to the full distribution let us briefly have a closer look at the transition probabilities \eqref{eq25}.
\begin{table}
\begin{tabular}{c|c|c|c|c}
$\ket{\nu}$ &  $\ket{0}$ & $\ket{1}$ & $\ket{2}$ & $\ket{3}$ \\ \hline\hline
$p(\ket{0}\rightarrow\ket{\nu})$ &  61.5$\%$ & 11.4$\%$ & 0.856$\%$ & 0.173$\%$
\end{tabular}
\caption{\label{tab:trans} \textbf{Transitions from bound state to bound state:} Transition probability \eqref{eq25} from one bound state to another for the Morse oscillator \eqref{eq16} driven by protocol \eqref{eq17} for parameters $\alpha=1$ and $t_f=5$ and with $V_0=1$, $\mu=1$, $k=10$ and $\hbar=1/2\pi$.}
\end{table}
Classical arguments let it appear plausible that the farther away (in energy) from the ground state a final state is located the less likely is a transition into this state. This is also reflected in the transition probabilities from one bound state to another, cf. Tab.~\ref{tab:trans}. 

\begin{figure}
\includegraphics[width = .46\textwidth]{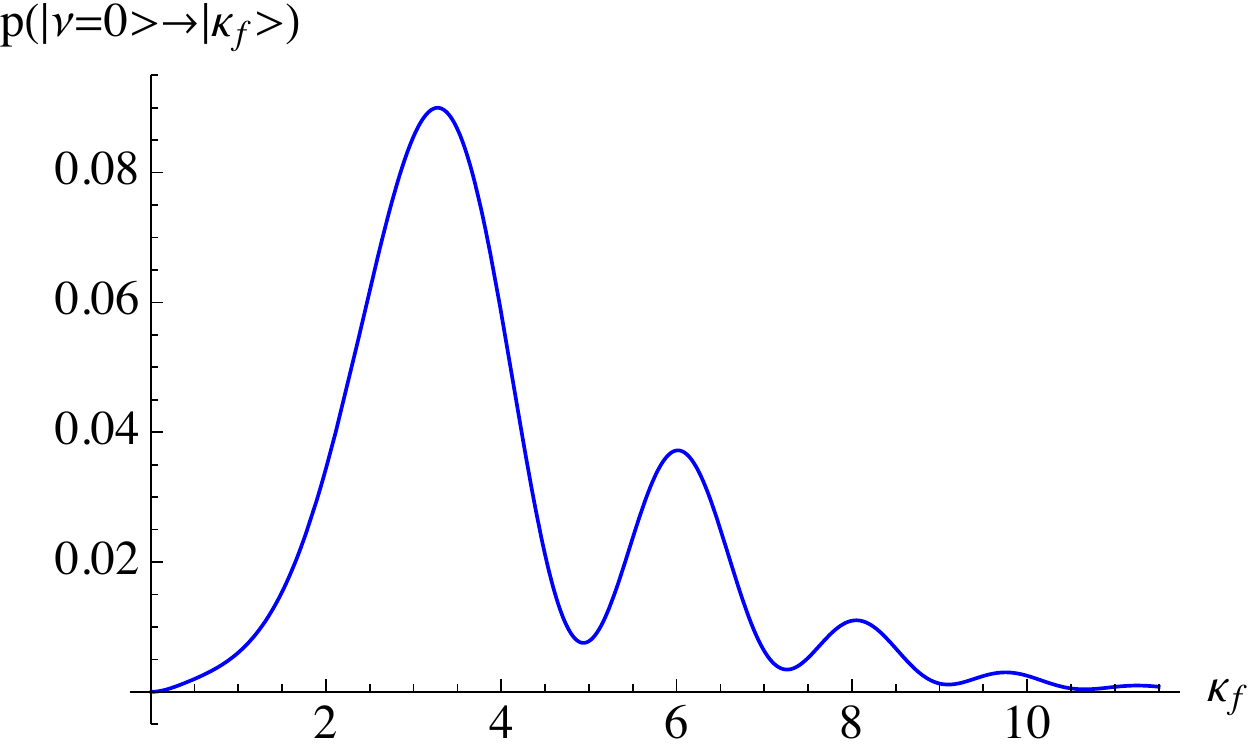}
\caption{\label{fig:trans} \textbf{Tunneling from ground state into scattering continuum:} Transition probability \eqref{eq25} from $\ket{\nu=0}$ into the scattering continuum for the Morse oscillator \eqref{eq16} driven by protocol \eqref{eq17} for parameters $\alpha=1$ and $t_f=5$ and with $V_0=1$, $\mu=1$, $k=10$ and $\hbar=1/2\pi$.}
\end{figure}

However, quantum mechanically this expectation is not entirely true. The transition probabilities 'measure' the overlap of initial and final state. Thus, classically unlikely or even forbidden transitions do occur. Comparing the plots in Figs.~\ref{fig:bound} and \ref{fig:scat} it becomes apparent that, for instance, the overlap of the ground state with an scattering state can be much larger than with any other bound state. This effect can be observed when plotting the transition probability from the ground state, $\ket{\nu=0}$, into the scattering continuum, $\ket{\kappa_f}$, during a scale invariant process \eqref{eq16}. We observe in Fig.~\ref{fig:trans} that tunneling \footnote{In the present context `tunneling' refers to energetically forbidden transitions in an  analogous, classical system.} into the scattering continuum is a non-negligible effect, which actually makes in this particular case about 26$\%$ of all transitions. We further observe that for specific values of $\kappa_f$ the transition probability \eqref{eq24} has local maximums. These 'dominant' transitions will be rediscovered in the structure of the work distribution shortly.

\section{\label{sec:work}Quantum work distribution}

In the previous sections we summarized properties of the Morse oscillator and discussed the solution of the dynamics for scale-invariant driving. In the remainder of the analysis we will now focus on the quantum work distribution \eqref{eq04}.

\subsection{Cumulative distribution function}

As we outlined earlier the resulting distribution function has two fundamentally different terms: a discrete part corresponding to transitions from the ground state to another bound state, and a continuous part corresponding to tunneling events from the ground state into the scattering continuum. 

To be able to illustrate these two contributions in a single plot, we consider the cumulative distribution,
\begin{equation}
\label{eq26}
\rho(W)=\int_{W} dW' \mathcal{P}(W')
\end{equation}
rather than the actual distribution function \eqref{eq04}. Hence, the discrete delta-peaks in $\mc{P}(W)$ yield a step function in $\rho(W)$, whereas the continuous contribution yields a continuous part.
\begin{figure}
\includegraphics[width = .46\textwidth]{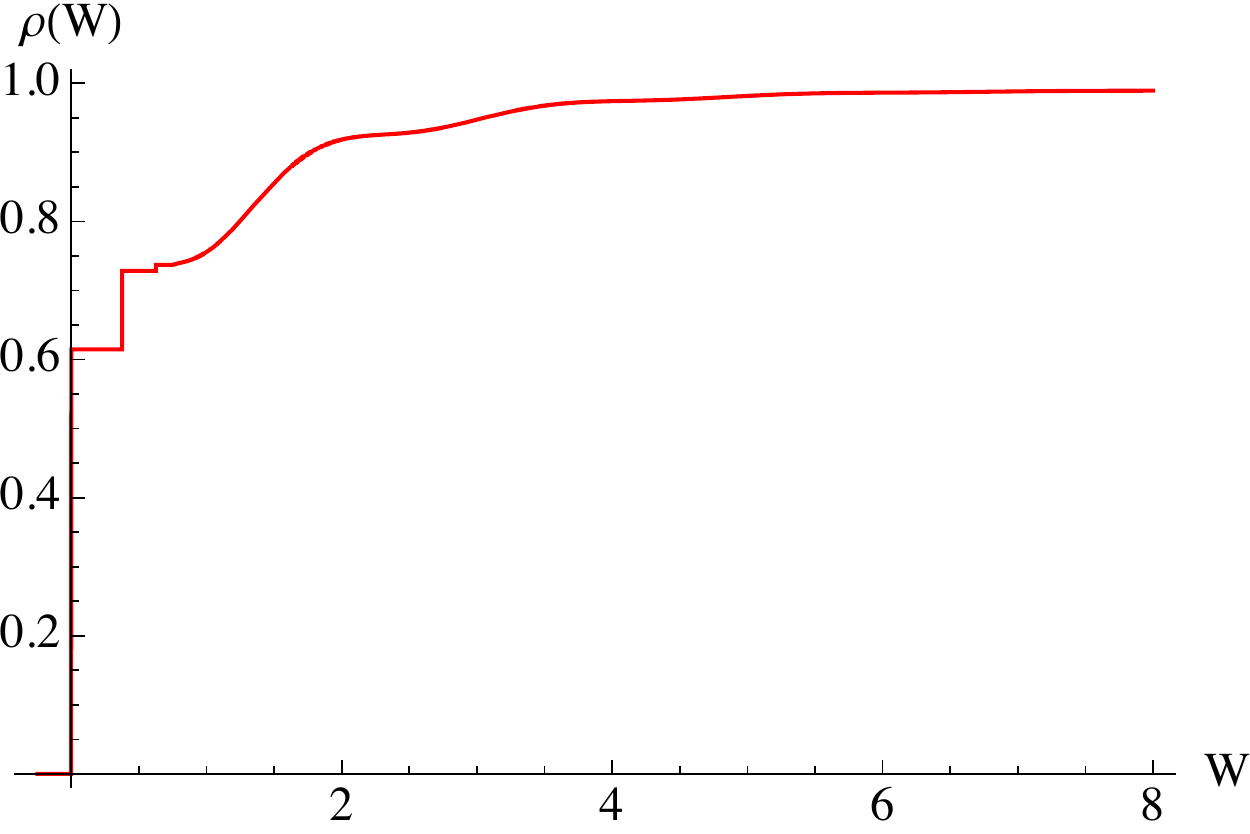}
\caption{\label{fig:dist}\textbf{Cumulative work distribution for the driven Morse oscillator:} Cumulative distribution function \eqref{eq26} for the scale-invariantly driven Morse oscillator \eqref{eq16} with protocol \eqref{eq17} for parameters $\alpha=1$, $t_f=5$, $V_0=1$, $\mu=1$, $k=10$ and $\hbar=1/2\pi$.}
\end{figure}

Figure~\ref{fig:dist} illustrates our final result, namely the cumulative work distribution \eqref{eq26} for the Morse oscillator under scale-invariant driving \eqref{eq16}. For small values of $W$, for the present parameters for $W\lesssim 1$, we observe the step function -- corresponding to transitions between bound states. However, we also observe that for $W\gtrsim 1$ the scattering states dominate the structure of the work distribution. Notice also the smooth 'humps' which correspond to the local maximums in the transition probability depicted in Fig.~\ref{fig:trans}.

\subsection{Exact results and the harmonic approximation}

\begin{figure}
\includegraphics[width = .46\textwidth]{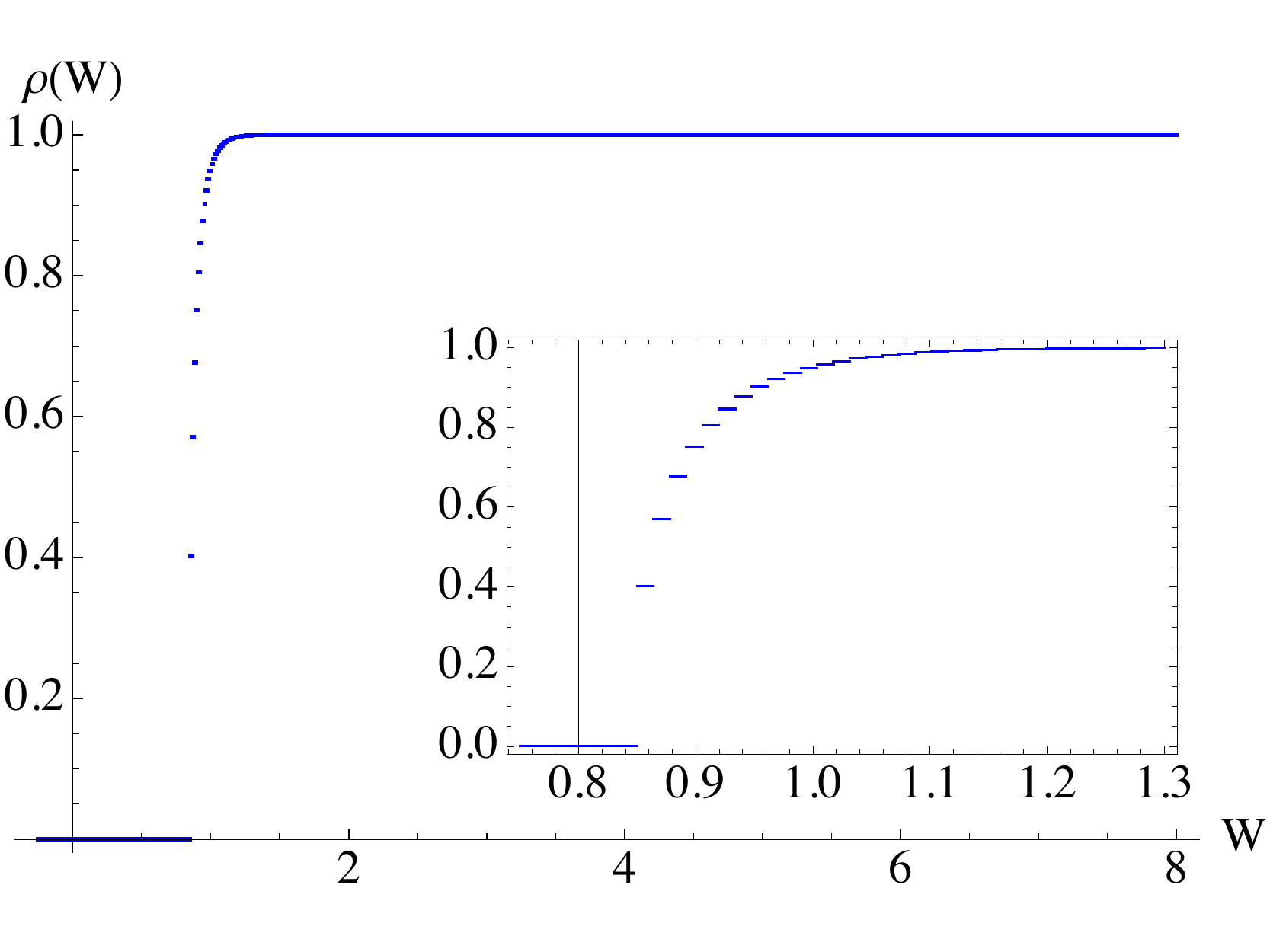}
\caption{\label{fig:comp}\textbf{Morse oscillator and harmonic approximation} Cumulative distribution function \eqref{eq26} for the scale-invariantly driven Morse oscillator \eqref{eq16} with protocol \eqref{eq17} in harmonic approximation for parameters $\alpha=1$, $t_f=5$, $V_0=1$, $\mu=1$, $k=10$ and $\hbar=1/2\pi$. The inset depicts a magnified version of the `interesting' part of the distribution.}
\end{figure}

We pointed out earlier that one commonly approximates the Morse oscillator by a harmonic oscillator. Thus, one might ask whether the work distribution can also be obtained by solving the harmonic problem, instead of going through the more involved analysis of the exact potential. To answer this question we compare the exact work distribution in Fig.~\ref{fig:dist} with the one that would be obtained by approximating the Morse potential \eqref{eq16} by a harmonic oscillator. For case of scale-invariant driving Eq.~\eqref{eq08} becomes,
\begin{equation}
\label{eq27}
V(x,t)\simeq -\frac{V_0}{\lambda_t^2}+\frac{V_0 B^2}{\lambda_t^4}\,x^2+\mc{O}(x^3)\,.
\end{equation}
For this system the quantum work distribution can be computed exactly, which has been extensively reported in Refs.~\cite{deffner_2008,deffner_2010}. Thus, we merely present the resulting plot in Fig.~\ref{fig:comp}.

We clearly observe that, at least for the here chosen parameters,  approximating the Morse potential by a harmonic oscillator is not justified. Whereas for the static case still some valid information might be obtained, the quantum work distribution and, more generally, the thermodynamic properties are quite different.

\section{\label{sec:con}Concluding Remarks}

In this paper we have outlined a method for determining the time-dependent wave functions of systems subject to scale-invariant driving. This method has been used to compute both transition probabilities and the quantum work distribution for the driven Morse oscillator. To the best of our knowledge, this analysis is the first to account for discrete and continuous parts of the eigenvalues spectrum and the consequences for the quantum work distribution. Even though we have been focused on the Morse potential as a case study, general insight can been obtained from our analysis as separation into bound and scattering states is a generic property of realistic systems. The presented method can be readily applied to any system for which the time-independent eigenfunctions are known. In this context Ref.~\cite{campo_2013,deffner_2014} presented a list of various potentials and their scale-invariant forms.

Thus, our method could also be applied to analyze the transition probabilities and work distributions for more complex quantum systems modeling, for instance, muscle fibers or proteins.

\subparagraph{Acknowledgements}
This work was realized under a Program of the Maryland Center for Undergraduate Research (MCUR). It is a pleasure to thank Zhiyue Lu, Christopher Jarzynski and Adolfo del Campo for stimulating discussions. SD acknowledges financial support by the National Science Foundation (USA) under grant DMR-1206971, and by the U.S. Department of Energy through a LANL Director's Funded Fellowship.

\appendix

\section{Discrete and continuous contributions}

In this appendix we summarize plots for the continuous and discrete parts of the quantum work distribution separately.

\subsection{Transitions bound-bound}

\begin{figure}
\includegraphics[width = .46\textwidth]{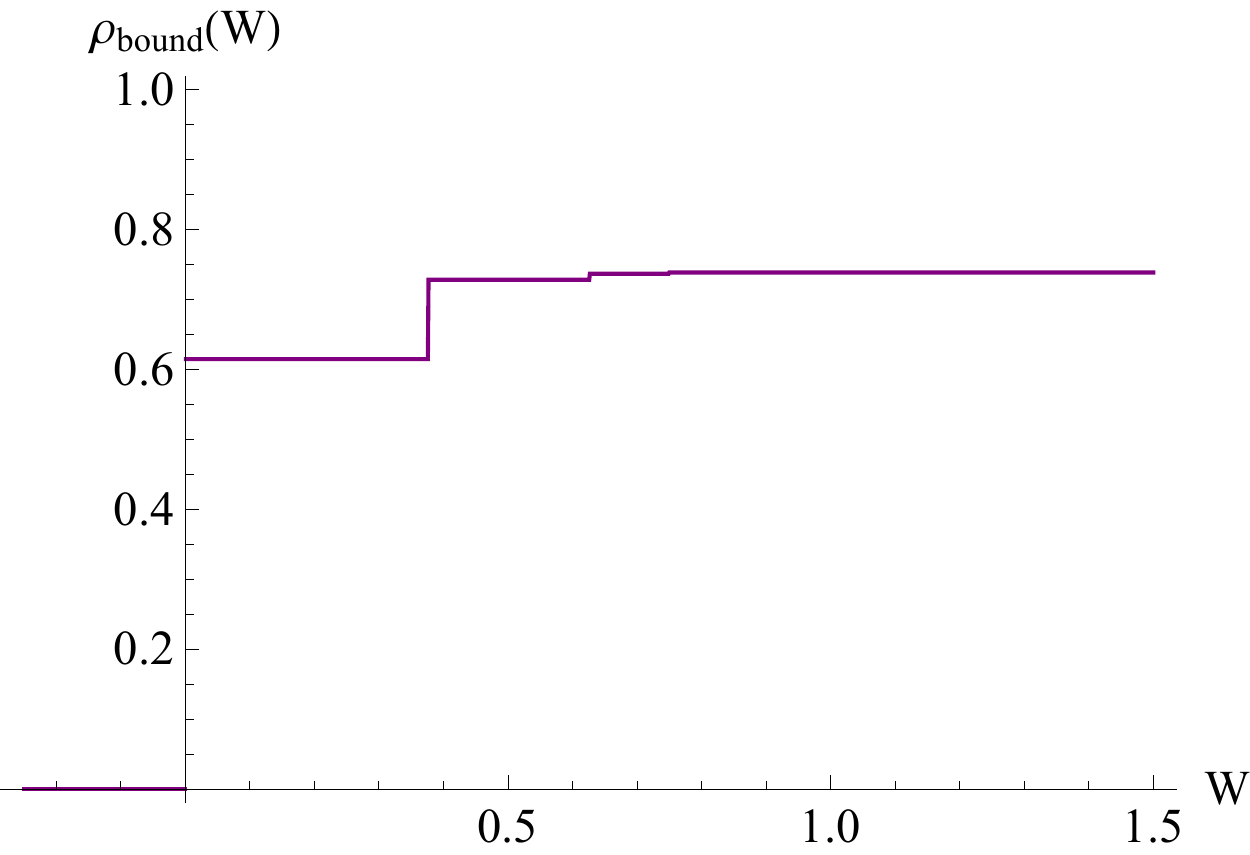}
\caption{\label{fig:dist_discrete}\textbf{Discrete part of the cumulative work distribution:} Discrete contribution to cumulative distribution function of Fig.~\ref{fig:dist} and accounting only for transitions between bound states.}
\end{figure}

In Fig.~\ref{fig:dist_discrete} we plot the discrete part of the cumulative distribution function in Fig.~\ref{fig:dist}. Observe that each `step' corresponds to a transition from the initial ground state $\ket{\nu=0}$ into another bound state.

\subsection{Transitions bound-scattering}

\begin{figure}
\includegraphics[width = .46\textwidth]{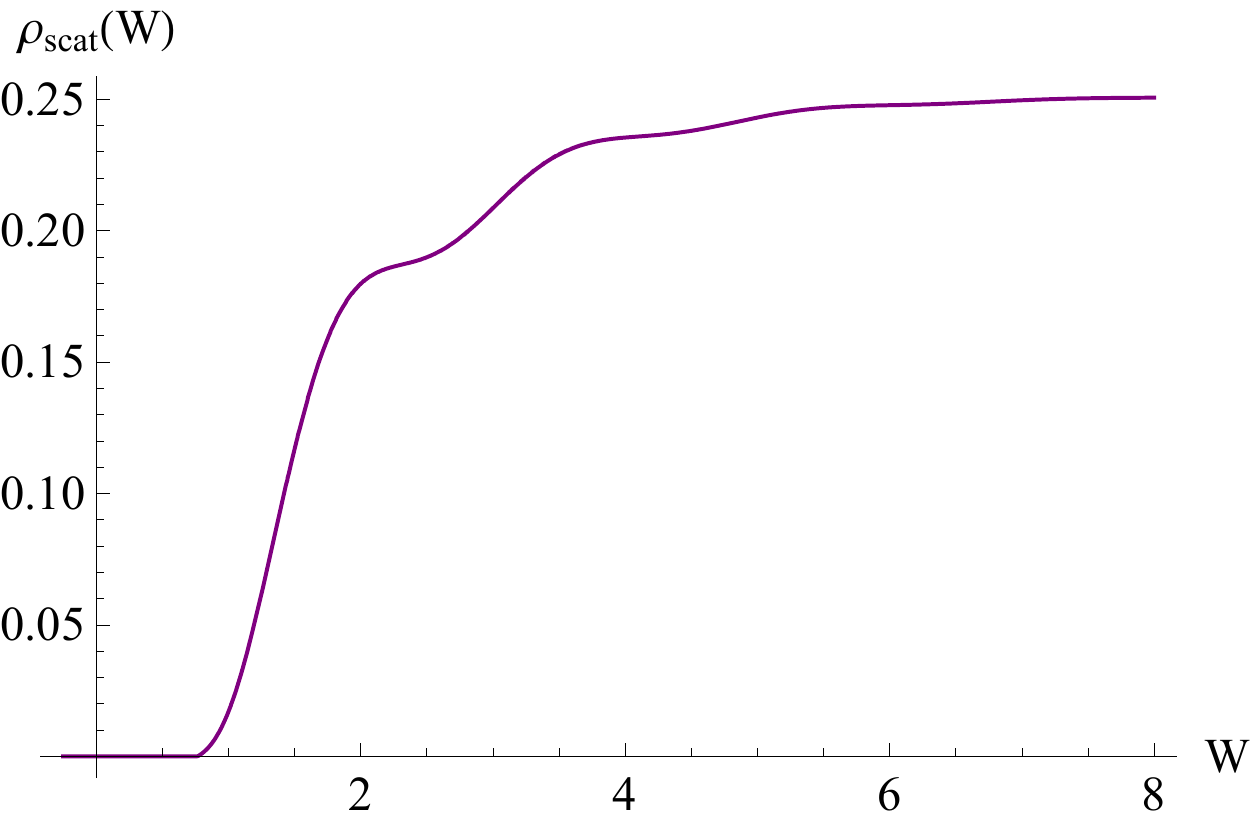}
\caption{\label{fig:dist_cont1}\textbf{Continuous part of the cumulative work distribution:} Continuous contribution to cumulative distribution function of Fig.~\ref{fig:dist} and accounting only for transitions between the initial ground state and the scattering continuum.}
\end{figure}

\begin{figure}
\includegraphics[width = .46\textwidth]{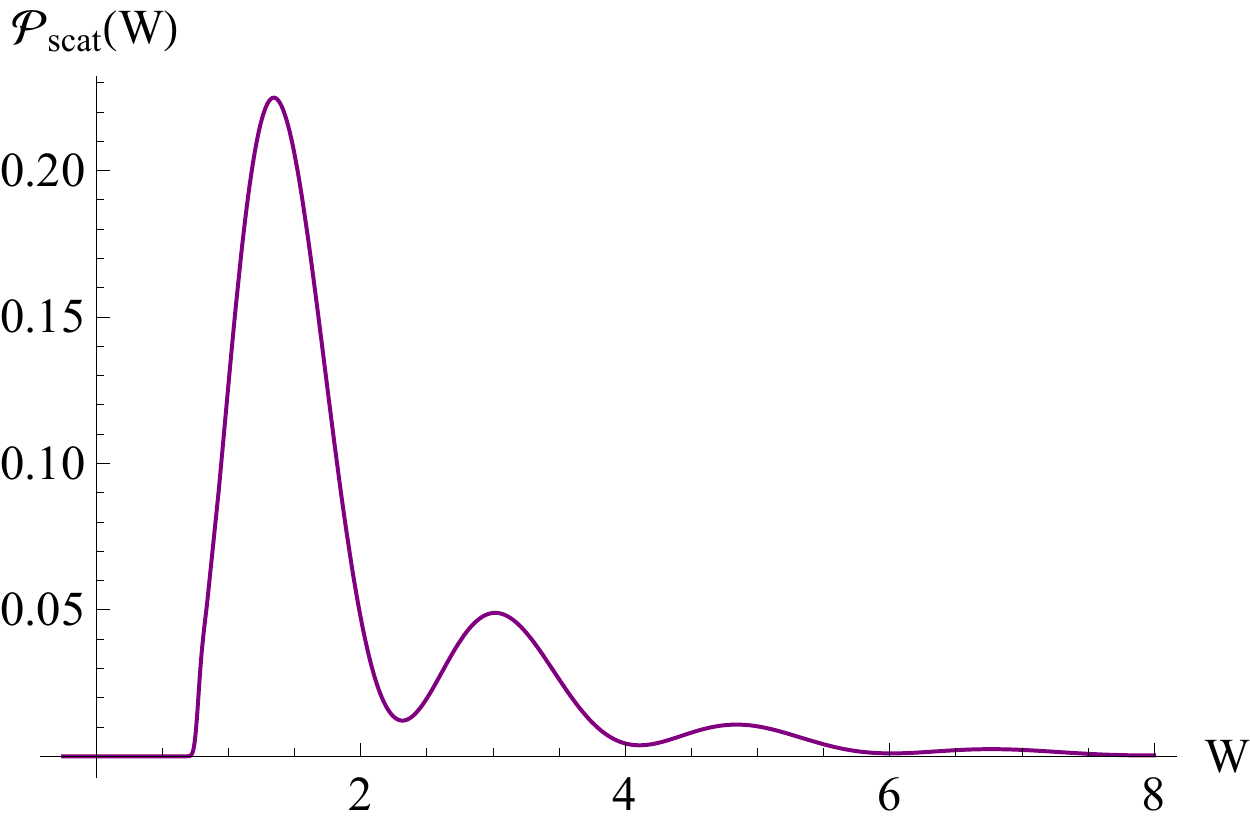}
\caption{\label{fig:dist_cont2}\textbf{Continuous part of the work probability density:} Continuous part of the probability density function \eqref{eq04} corresponding to Fig.~\ref{fig:dist} and accounting only for transitions between initial ground state and the scattering continuum.}
\end{figure}

Finally, Figs.~\ref{fig:dist_cont1} and \ref{fig:dist_cont2} illustrate the continuous part of the distribution in Fig.~\ref{fig:dist}. To highlight the effect of the `resonances', see also Fig.~\ref{fig:trans}, we plot the cumulative distribution \eqref{eq26} in Fig.~\ref{fig:dist_cont1} and the probability density \eqref{eq04} in Fig.~\ref{fig:dist_cont2}.
Notice that each `hump' in Fig.~\ref{fig:dist_cont2} corresponds to a resonance, i.e., local maximum of the continuous transitions probabilities \eqref{eq24}.


\begin{thebibliography}{10}
\expandafter\ifx\csname url\endcsname\relax
  \def\url#1{\texttt{#1}}\fi
\expandafter\ifx\csname urlprefix\endcsname\relax\def\urlprefix{URL }\fi
\expandafter\ifx\csname href\endcsname\relax
  \def\href#1#2{#2} \def\path#1{#1}\fi

\bibitem{kurchan_2000}
J.~Kurchan, {A Quantum Fluctuation Theorem}, arXiv cond-mat/0007360.

\bibitem{tasaki_2000}
H.~Tasaki, {Jarzynski Relations for Quantum Systems and Some Applications},
  arXiv condmar/0009244.

\bibitem{talkner_2007}
P.~Talkner, E.~Lutz, P.~H\"anggi, {Fluctuation theorems: Work is not an
  observable}, \PRE 75 (2007) 50102.

\bibitem{campisi_2011}
M.~Campisi, P.~H\"anggi, P.~Talkner, {Colloquium: Quantum fluctuation
  relations: Foundations and applications}, \RMP 83 (2011) 771.

\bibitem{deffner_2008}
S.~Deffner, E.~Lutz, {Nonequilibrium work distribution of a quantum harmonic
  oscillator}, \PRE 77 (2008) 021128.

\bibitem{talkner_2008}
P.~Talkner, P.~S. Burada, P.~H\"anggi, {Statistics of work performed on a
  forced quantum oscillator}, \PRE 78 (2008) 11115.

\bibitem{talkner_2008a}
P.~Talkner, P.~H\"anggi, M.~Morillo, {Microcanonical quantum fluctuation
  theorems}, \PRE 77 (2008) 051131.

\bibitem{deffner_2010_prl}
S.~Deffner, E.~Lutz, {Generalized Clausius Inequality for Nonequilibrium
  Quantum Processes}, \PRL 105.

\bibitem{deffner_2010}
S.~Deffner, O.~Abah, E.~Lutz, Quantum work statistics of linear and nonlinear
  parametric oscillators, Chem. Phys. 375 (2010) 200.

\bibitem{deffner_2011}
S.~Deffner, E.~Lutz, {Nonequilibrium Entropy Production for Open Quantum
  Systems}, \PRL 107.

\bibitem{quan_2012}
H.~Quan, C.~Jarzynski, {Validity of nonequilibrium work relations for the
  rapidly expanding quantum piston}, \PRE 85 (2012) 031102.

\bibitem{fusco_2014}
L.~Fusco, S.~Pigeon, T.~J.~G. Apollaro, A.~Xuereb, L.~Mazzola, M.~Campisi,
  A.~Ferraro, M.~Paternostro, G.~{De Chiara}, {Assessing the non-equilibrium
  thermodynamics in a quenched quantum many-body system via single projective
  measurements}, \PRX 4 (2014) 031029.

\bibitem{huber_2008}
G.~Huber, F.~Schmidt-Kaler, S.~Deffner, E.~Lutz, {Employing Trapped Cold Ions
  to Verify the Quantum Jarzynski Equality}, \PRL 101 (2008) 070403.

\bibitem{mazzola_2013a}
L.~Mazzola, G.~{De Chiara}, M.~Paternostro, {Measuring the Characteristic
  Function of the Work Distribution}, \PRL 110 (2013) 230602.

\bibitem{dorner_2013}
R.~Dorner, S.~R. Clark, L.~Heaney, R.~Fazio, J.~Goold, V.~Vedral, {Extracting
  Quantum Work Statistics and Fluctuation Theorems by Single-Qubit
  Interferometry}, \PRL 110 (2013) 230601.
  
\bibitem{batahlao_2014}
T.~B.~Batalh\~ao, A.~M.~Souza, L.~Mazzola, R.~Auccaise, R.~S.~Sarthour, I.~S.~Oliveira, J.~Goold, G.~{De Chiara}, M.~Paternostro, R.~M.~Serra, Roberto,
  {Experimental Reconstruction of Work Distribution and Study of Fluctuation Relations in a Closed Quantum System}, \PRL 113    (2014) 140601.

\bibitem{abah_2012}
O.~Abah, J.~Ro\ss~nagel, G.~Jacob, S.~Deffner, F.~Schmidt-Kaler, K.~Singer,
  E.~Lutz, Single-ion heat engine at maximum power, \PRL 109 (2012) 203006.

\bibitem{abah_2013}
O.~Abah, E.~Lutz, {Efficiency of heat engines coupled to nonequilibrium
  reservoirs}, EPL (Europhysics Letters) 106 (2013) 20001.

\bibitem{rossnagel_2013}
J.~Ro\ss~nagel, O.~Abah, F.~Schmidt-Kaler, K.~Singer, E.~Lutz, {A nano heat
  engine beyond the Carnot limit}, \PRL 112 (2014) 030602.

\bibitem{campo_2014}
A.~del Campo, J.~Goold, M.~Paternostro, More bang for your buck: Towards
  super-adiabatic quantum engines, Sci. Rep. 4 (2014) 6208.

\bibitem{deffner_2014}
S.~Deffner, C.~Jarzynski, A.~del Campo, {Classical and quantum shortcuts to
  adiabaticity for scale-invariant driving}, \PRX 4 (2014) 021013.

\bibitem{zheng_2014}
Y.~Zheng, D.~Poletti, Work and efficiency of quantum otto cycles in power-law
  trapping potentials, Phys. Rev. E 90 (2014) 012145.

\bibitem{berry_1984}
M.~V. Berry, G.~Klein, {Newtonian trajectories and quantum waves in expanding
  force fields}, J. Phys. A: Math. Gen. 17 (1984) 1805.

\bibitem{alhassid_1983}
Y.~Alhassid, F.~Iachello, F.~G\"{u}rsey, {Group theory of the Morse
  oscillator}, Chem. Phys. Lett. 99 (1983) 27.

\bibitem{alhassid_1984}
Y.~Alhassid, {Algebraic calculation of the Morse oscillator scattering matrix},
  Chem. Phys. Lett. 108 (1984) 79.

\bibitem{benjamin_1985}
I.~Benjamin, R.~H. Bisseling, R.~Kosloff, R.~D. Levine, J.~Manz, H.~H.~R.
  Schor, {Quasi-bound states of coupled Morse oscillators}, Chem. Phys. Lett.
  116~(4) (1985) 255.

\bibitem{dahl_1988}
J.~P. Dahl, M.~Springborg, {The Morse oscillator in position space, momentum
  space, and phase space}, J. Chem. Phys. 88 (1988) 4535.

\bibitem{frank_2000}
A.~Frank, A.~Rivera, K.~Wolf, {Wigner function of Morse potential eigenstates},
  \PRA 61 (2000) 054102.

\bibitem{dong_2002}
S.-H. Dong, R.~Lemus, A.~Frank, {Ladder operators for the Morse potential},
  Int. J. Quan. Chem. 86 (2002) 433.

\bibitem{rawitscher_2002}
G.~Rawitscher, C.~Merow, M.~Nguyen, I.~Simbotin, {Resonances and quantum
  scattering for the Morse potential as a barrier}, Am. J. Phys. 70 (2002) 935.

\bibitem{demirplak_2005}
M.~Demirplak, S.~A. Rice, {Assisted adiabatic passage revisited}, J. Phys.
  Chem. B 109 (2005) 6838.

\bibitem{angelova_2008}
M.~Angelova, V.~Hussin, {Generalized and Gaussian coherent states for the Morse
  potential}, J. Phys. A: Math. Theor. 41 (2008) 304016.

\bibitem{dittrich_2010}
T.~Dittrich, E.~A. G\'{o}mez, L.~A. Pach\'{o}n, {Semiclassical propagation of
  Wigner functions}, J. Chem. Phys. 132 (2010) 214102.

\bibitem{mccoy_2011}
A.~B. McCoy, {Curious properties of the Morse oscillator}, Chem. Phys. Lett.
  501 (2011) 603.

\bibitem{morse_1929}
P.~M. Morse, {Diatomic molecules according to the wave mechanics. II.
  Vibrational levels}, \PR 34 (1929) 57.

\bibitem{flugge_1971}
S.~Fl\"ugge, Practical quantum mechanics, Vol.~I, Springer, Heidelberg,
  Germany, 1971.

\bibitem{kondo_1988}
A.~E. Kondo, D.~R. Truax, {On a time-dependent extension of the Morse
  potential}, J. Math. Phys. 29 (1988) 1396.

\bibitem{lima_2006}
E.~F. de~Lima, J.~E.~M. Hornos, {The Morse oscillator under time-dependent
  external fields}, J. Chem. Phys. 125 (2006) 164110.

\bibitem{kafri_2012}
D.~Kafri, S.~Deffner, {Holevo’s} bound from a general quantum fluctuation
  theorem, \PRA 86 (2012) 044302.

\bibitem{deffner_2013}
S.~Deffner, {Quantum entropy production in phase space}, EPL (Europhysics
  Letters) 103 (2013) 30001.

\bibitem{kaplan_2003}
I.~G. Kaplan, {Handbook of Molecular Physics and Quantum Chemistry}, Wiley, New
  York City, NY, USA, 2003.

\bibitem{abramowitz_1964}
M.~Abramowitz, I.~A. Stegun, {Handbook of mathematical functions with formulas,
  graphs, and mathematical tables}, Washington D.C., USA, 1964.

\bibitem{jarzynski_2013}
C.~Jarzynski, Generating shortcuts to adiabaticity in quantum and classical
  dynamics, \PRA 88 (2013) 040101.

\bibitem{campo_2013}
A.~del Campo, Shortcuts to adiabaticity by counterdiabatic driving, \PRL 111
  (2013) 100502.

\end{thebibliography}

\end{document}